\documentclass[a4paper,fleqn,usenatbib]{mnras}

\usepackage[T1]{fontenc}
\usepackage{ae,aecompl}

\usepackage{graphicx}	
\usepackage{amsmath}	
\usepackage{amssymb}	

\title[Morpho-Photometric Redshifts]{Morpho-Photometric Redshifts}

\author[K. Menou]{
Kristen Menou,$^{1,2}$
\\
$^{1}$  Physics \& Astrophysics Group, Dept.   of  Physical  \&  Environmental  Sciences,  University  of  Toronto  Scarborough,\\   1265 Military Trail, Toronto, Ontario, M1C 1A4, Canada \\
$^{2}$ Dept.  of Astronomy \& Astrophysics, University of Toronto.
50 St.  George Street, Toronto, Ontario, M5S 3H4, Canada \\
}

\date{Accepted XXX. Received YYY; in original form ZZZ}

\pubyear{2017}

\begin{document}
\label{firstpage}
\pagerange{\pageref{firstpage}--\pageref{lastpage}}
\maketitle

\begin{abstract}
{Machine learning (ML) is one of two standard approaches (together with SED fitting) for estimating the
redshifts of galaxies when only photometric information is
available.} ML photo-z solutions have traditionally ignored the
morphological information available in galaxy images or partly
included it in the form of hand-crafted features, with mixed results. We train a
morphology-aware photometric redshift machine using modern deep
learning tools. It uses a custom architecture that jointly trains on galaxy
fluxes, colors and images. Galaxy-integrated quantities are
fed to a Multi-Layer Perceptron (MLP) branch while images are
fed to a convolutional (convnet) branch that can
learn relevant morphological features. This split MLP-convnet
architecture, which aims to disentangle strong photometric features from comparatively weak morphological ones, proves important for strong
performance: a regular convnet-only architecture, while exposed to all available photometric
information in images, delivers comparatively poor performance. 
We present a cross-validated MLP-convnet model trained on 130,000
SDSS-DR12 galaxies that outperforms a
hyperoptimized Gradient Boosting solution (hyperopt+XGBoost), as well as the equivalent MLP-only architecture, on the redshift bias metric. The 4-fold cross-validated MLP-convnet model achieves a bias $\delta z / (1+z) =-0.70 \pm  1 \times 10^{-3} $, approaching the performance of a reference ANNZ2 ensemble of 100 distinct models trained on a comparable dataset. The relative performance of the morphology-aware and morphology-blind models indicates that galaxy morphology does improve {ML-based} photometric redshift estimation. 
\end{abstract}






\section{Introduction}

Spectroscopic redshits of galaxies are used as distance measures to study large-scale structure in the Universe. It has always been challenging to obtain the spectra, and thus reliably measure the redshifts, of the numerous galaxies imaged on the sky by astronomers, because of practical limitations in the resources that can be allocated to spectroscopic follow-up surveys. As an example, SDSS has imaged roughly 200 million galaxies while it obtained the spectra of about 2 million of those \citep{2016MNRAS.460.1371B}. As we approach the era of the Large Synoptic Sky Survey (LSST), with vast dedicated imaging capabilities but limited systematic spectroscopic follow-up, the  challenge of measuring the redshift of imaged galaxies has only become more acute \citep{2018AJ....155....1G, 2018NatAs.tmp...68S}.

Machines trained to estimate galaxy redshifts based on their integrated fluxes/colors are one of the two main approaches being considered to tackle the challenge of photometric redshift ("photo-z") estimations. A variety of algorithms have been applied and trained on this problem in the supervised learning framework, whereby the subset of imaged galaxies for which spectroscopic redshifts have been measured are used as a training set. Post-training, the redshifts of the vast majority of galaxies, without spectra, are estimated by the trained machine, based on photometric data only.

In an attempt to improve photo-z estimations, various studies have considered adding morphological information, beyond the usual galaxy-integrated colors, as training input for the machine.\footnote{Morphological information is used in the SED-fitting approach to  photometric redshifts, e.g. as a prior \citep[e.g.][]{2000ApJ...536..571B}} Results have been somewhat inconclusive, in the sense that performance gains are not systematic when morphological data is added:  performance gains/losses from morphological data depend on details of the task/dataset/method adopted \citep[see][and references therein for a detailed account]{2018MNRAS.475.3613S}. {Reports of gains or losses depending on which combination of morphological features are included are not entirely surprising:} feature engineering is a notoriously 'non-linear' process\footnote{For example, two features that are found to beneficial in isolation have no guarantee to remain beneficial if used together.} and optimizing over the feature space for best performance is a generally complex enterprise \citep[e.g.,][]{2018A&A...616A..97D}.

Here, we present a machine learning (ML) focused study of photo-z estimation that attempts to clarify the role that morphology has in promoting better redshift estimations. Our approach is to use the modern tools of deep learning,  in the form of a deep convolutional network (convnet). One of the key achievements in computer vision over the past few years has been the demonstration that properly trained convnets can automatically learn powerful and generizable features for a variety of tasks on image data \citep[see Chapter 9,][]{Goodfellow-et-al-2016}. Therefore, the promise of deep learning in the photo-z context is an ability to build a machine that automatically learns strong morphological features for photo-z estimation. {While convnets have been applied to the photo-z estimation task before \citep{2016A&C....16...34H, 2018A&A...609A.111D}, this approach is still fairly new and the improvements one can expect have not been fully clarified \citep{2018NatAs.tmp...68S}. In this work, we report clear improvements in photo-z estimation performance on the redshift bias metrics} with the inclusion of morphological data, but only when a custom network architecture is adopted that treats separately galaxy-integrated features and the image data containing morphological information.

In Section~2, we describe our dataset acquisition and processing pipeline. In Section~3, we detail our machine learning approach. Results can be found in Section~4, before we conclude in Section~5. 


\section{Dataset Assembly and Preparation}

\subsection{Data Acquisition}
To assemble the dataset, we closely follow the methodology detailed in \cite{2016A&C....16...34H}. A catalog of SDSS-DR10 galaxies with spectroscopically-measured redshifts is obtained by executing the same SQL command on the SDSS CasJob server as described in the Appendix of \cite{2016A&C....16...34H}. One additional variable request is made for the u-band extinction data. This results in a spectroscopic catalog of approximately 1.92M galaxies with {\it ugriz} photometric information and measured spectroscopic redshifts.

To build a training/validation set, 5-band images are downloaded for about 136000 galaxies listed in our spectroscopic catalog. A distinct test set is built by downloading an additional 26000 images (making sure there is no overlap with the training set). All image downloads are from SDSS-DR12 data. When multiple objects/spectra exist for a given catalog entry, only coordinates for the first listed entry is used to avoid duplicates. The {\tt astropy} SDSS query module is used for plate downloads. After download, 
cut-outs of 72x72 pixels centered on the galaxy catalog coordinates of interest are extracted in 5 bands and stored for training/validation/testing (larger 144x144 cutouts were also experimented with and eventually discarded). Small cut-outs are computationally efficient at training time and they guarantee that the vast majority of images collected are images of galaxies in isolation, with only limited information on the larger scale galactic environment.

{With an SDSS pixel size of $0.396$", our  72x72 cutouts corresponds to an angular size of $28.5$". The size of these cutouts are, in some sense, tuned to the density of sources on the sky, which depends on the depth of the catalogue of interest. On the other hand, at redshifts $z<0.1$, a 50kpc-sized galaxy starts overfilling the 72x72 cutouts. As a result, morphological information is effectively truncated for the largest catalog sources,  at low enough redshifts. About 25\% of our sources lie at  $z< 0.1$ (see Fig.~2) so this limitation could have a sizable impact on our models using image cutouts. We did not investigate further this limitation but note that the consistency of our results using larger 144x144 cutouts suggests that the impact is relatively minor overall (though low redshift estimates may be biased).}

{To summarize, all raw fluxes (and $r$-band Petrosian radii) are taken from the SDSS DR10 catalogs while raw images are direct cutouts from SDSS DR12 plates.
All fluxes and images are then corrected for extinction. }We follow the method of \cite{2017MNRAS.464.4463K} for SDSS images, by first converting images from fluxes to luptitudes, applying extinction corrections then transforming back to fluxes.\footnote{ {\tt https://github.com/EdwardJKim/dl4astro}}

\subsection{Data Preparation}

{We experimented with various pre-processing steps on the raw input data, since it is not a priori clear which processed format yields the best performance at training time.} Performance on the validation set was used to settle on the adopted input format. Color cuts were applied to improve sample quality, following the suggested color cuts  of \cite{2016MNRAS.460.1371B}:
\begin{equation}
\begin{aligned}
	-0.911 < (u-g) &< 5.597 \\
	0.167 < (g-r) &< 2.483 \\
	0.029 < (r-i) &< 1.369 \\
	-0.452 < (i-z) &< 0.790 \\
\end{aligned}
\end{equation}
No photometric error cut was applied however \citep[unlike][]{2016MNRAS.460.1371B}.

\begin{figure}
	\includegraphics[width=9cm]{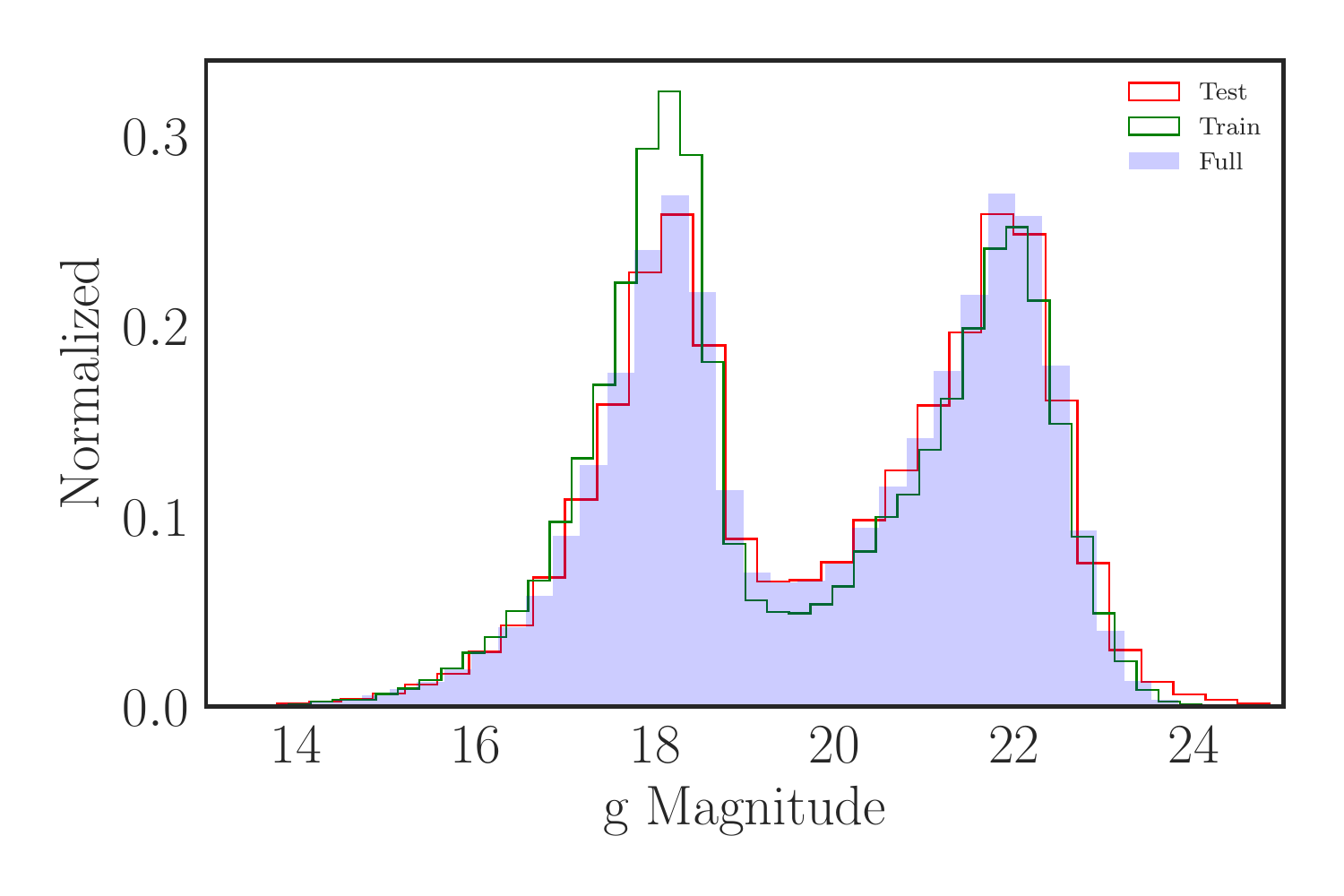}
    \caption{Distribution of $g$ magnitude in the train (green), test (red) and full (filled blue, representing all 1.92 million samples) datasets. Histograms have been normalized for ease of comparison.}
    \label{fig:example_figure}
\end{figure}

\begin{figure}
	\includegraphics[width=9cm]{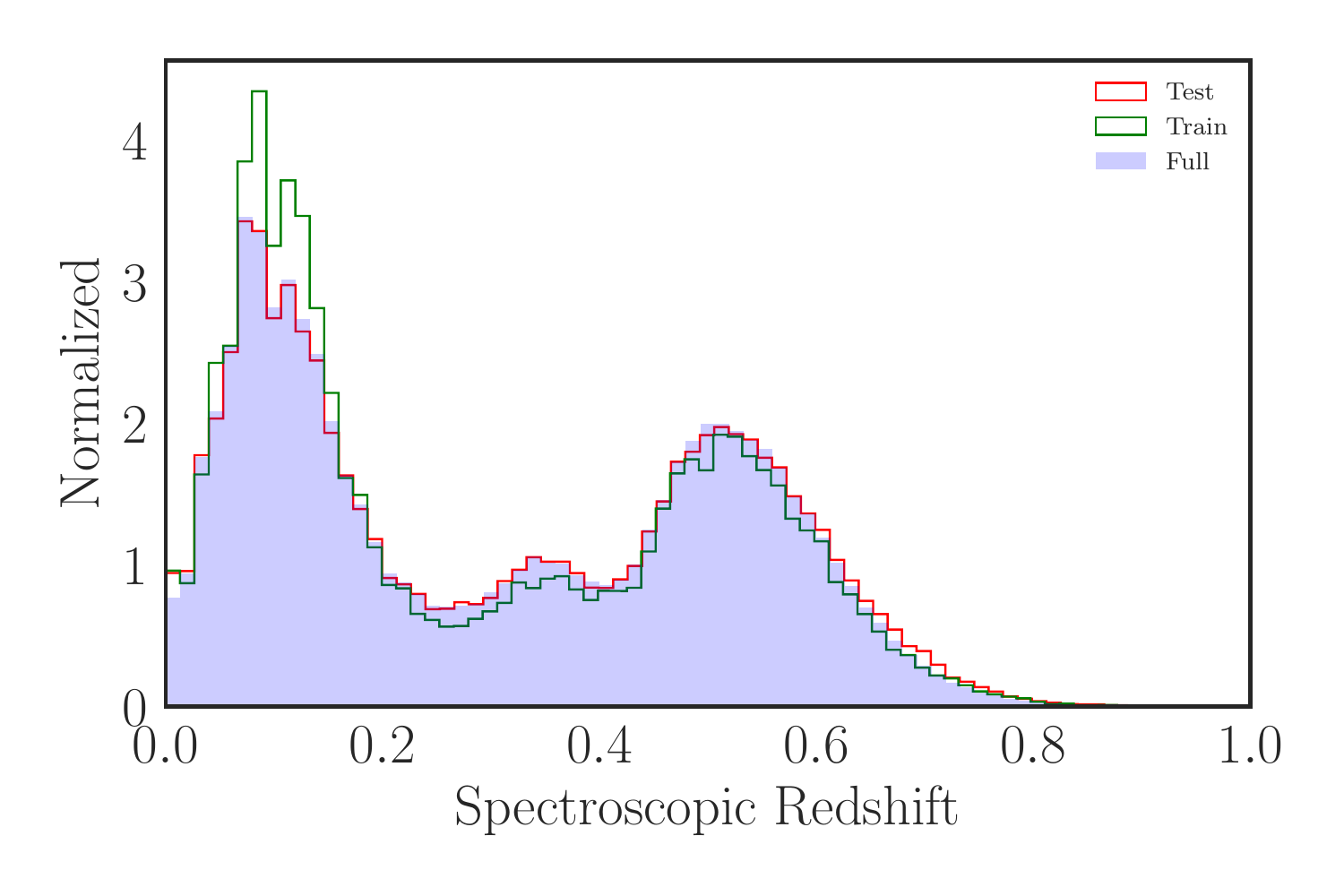}
    \caption{Same as Figure 1, showing the train/test/full dataset distributions in spectroscopic redshift. Histograms have been normalized for ease of comparison.}
    \label{fig:example_figure}
\end{figure}

Scalar input data is composed of 10 features: the 5 SDSS magnitudes, 4 SDSS colors and the $r$-band Petrosian radius. All these permutation-invariant\footnote{The specific order in which these features are fed to the MLP branch is unimportant (as long as it remains fixed). By contrast, pixel features in images are fed to the convnet in a way that guarantees spatial continuity.} features are normalized to have zero-mean and unit-variance using the {\tt scikit-learn} StandardScaler built-in function. The scaler is fit on the training set and then applied to the test set. Image data is prepared in flux units, rather than magnitudes, over 5 channels: 4 flux difference channels ($F_u-F_g$, $F_g-F_r$, $F_r-F_i$, $F_i-F_z$), and a 'stacked' flux which is the sum of the all the {\it ugriz} flux images. Note that the flux difference images differ from the colors traditionally used by astronomers (which are logarithmic flux ratio). The flux images are then simply normalized, per galaxy, to the max pixel value across all channels for that galaxy. This normalization guarantees that galaxy-specific spectral information remains encoded in the four flux-difference channels.

{By construction, our training and test sets have distributions that are very similar to each other and to those of the full catalogue dataset. This is illustrated in Figs~1 and~2 where the distributions of each dataset in $g$ magnitude and in spectroscopic redshift space are compared. Note that each dataset extends to $z \simeq 2$ but samples beyond  $z \simeq 0.8$ as so rare as to be inconsequential.}

\section{Model Architecture and Training}

\subsection{Mixed MLP-Convnet Architecture}

The main idea leading to the network architecture presented in this paper originates from the well-documented astrophysical knowledge that photometric (galaxy-integrated) colors are the primary variables correlating with the spectroscopic redshifts of galaxies, while galaxy morphology is typically of secondary importance \citep{2018MNRAS.475.3613S}. This means that a convnet presented with images (only) will first need to learn to integrate the fluxes of galaxy over their spatial extent to perform well on the redshift regression task before it can focus on secondary morphological details. Conversely, a MLP (or any other standard ML algorithm) can do well on the photo-z estimation task, once presented with the most meaningful galaxy-integrated colors as input features.

Given this, one way to help a network use photometry effectively and still learn additional/secondary features from morphological details in images is to feed both types of information to the network through two separate branches. Optimizing such a split network on the full redshift regression task should permit the complementarity of photometric and morphological information to be used optimally as backpropagating gradients flow through each network branch using all input features as best as it can \cite[see Chapter 6,][]{Goodfellow-et-al-2016}.

This observation naturally leads to the concept of a mixed MLP-convnet architecture with concatenated features, since the MLP branch should perform well on photometric quantities and the convnet should do well on complementary morphogical information. Ideally, the convnet will learn the best complementary features to the photometric quantities, and in so could outperform the hand-crafted morphological features that have been traditionally used by astronomers \citep[][and references therein]{2018MNRAS.475.3613S}.

The final architecture adopted in this work is composed of two separate branches: an MLP branch  with 10-feature input and a convnet branch with 5-channel 72x72 image input. Details of the architectures are provided in Table~1. Our models were built with {\tt Keras} on top of a {\tt Tensorflow} backend and we use keras-specific terminology to describe our network. The MLP branch has two hidden layers with 4 fully-connected (dense) neurons each. PReLU activations are used in the MLP branch. The convnet branch has 4 convolutional blocks (with 2 convolutional layers each), supplemented with a maxpooling operation at the end of the first three blocks. The convolutional features are collected at the very end of the convnet branch via a global average-pooling operation and subsequently concatenated with the MLP branch features through a Merge layer, before the final regression layer. Note that we pose the problem as a regression task, rather than a 100-bin multi-class classification task like \cite{2016A&C....16...34H}.

All convolutional layers use 3x3 kernels, followed by ReLU activations. Filter sizes are doubled at the end of each convolutional block (16->32->64->128). MaxPooling2D operates with (2,2) strides. All layers are initialized with an 'he\_normal' weight distribution. This convnet architecture is inspired by the seminal VGG network architecture designed for Imagenet \citep{Simonyan14c}. No extra regularization, batch-normalization or drop-out layers were used, for simplicity. Note that we use a relatively small number of filters relative to standard ImageNet-scale architectures \citep{Simonyan14c}.
  
Some degree of architecture optimization (monitoring the validation performance) was performed on the convnet branch, specifically in terms of filter size, kernel size and number of convolutional blocks, but there is likely still room for improvement using more modern architectures/blocks than the simple VGG scheme. The MLP architecture was not optimized, with the exception of activations (PReLU seems to perform somewhat better than ReLU). 

With the mixed MLP-convnet architecture adopted here, the model has the potential to learn all there is to learn in the galaxy-integrated quantities fed to the MLP branch, leaving the convnet branch free to learn the best complementary features that might exist in images, in particular all relevant morphological features (beyond the r-band Petrosian radius that is explicitly fed to the MLP branch). The split network architecture also allows us to test our main hypothesis on the role of morphological features, by selectively zeroing the input of the convnet branch (or the MLP branch) in order to compare the contribution of each branch in isolation vs together. 
In what follows, we refer to these modified models as MLP-only and convnet-only, respectively. One of the main goals of our work is to find out whether a mixed  MLP-convnet model can indeed outperform the equivalent MLP-only model, which would indicate that morphogical features can indeed be efficiently learned to provide beneficial returns to photo-z estimation.

\begin{table}
	\centering
	\caption{Mixed MLP-convnet architecture proposed in this work. Images (72x72) are fed to a traditional convnet (left) while 10 permutation-invariant  features are fed to an MLP branch (right). The representations learned by each branch are concatenated and used for the redshift estimation in the form of  a regression task. All convolution layers use 3x3 kernels followed by ReLU activations.  }
	\label{tab:example_table}
	\begin{tabular}{cc} 
		\hline
		Convnet Branch & MLP Branch \\
		\hline
		        InputLayer (72, 72, 5)         & \\                                      

              Conv2D (72, 72, 16)                     & \\                          

              Conv2D (72, 72, 16)                 &\\                              

        MaxPooling2D (36, 36, 16)          & \\                                     

              Conv2D (36, 36, 32)             & \\                                  

              Conv2D (36, 36, 32)            & \\                                   

        MaxPooling2D (18, 18, 32)         &\\                                      

              Conv2D (18, 18, 64)            & \\                                   

              Conv2D (18, 18, 64)         &             InputLayer (10)    \\            

        MaxPooling2D (9, 9, 64)         & Dense (4)  \\                

              Conv2D (9, 9, 128)            & PReLU (4) \\  
                    
              Conv2D (9, 9, 128)            & Dense (4)   \\              

GlobalAveragePooling2D (128)          & PReLU (4)  \\   
	\end{tabular}
	\begin{tabular}{c} 
	Merge (132)  \\                             
    Dense (1) \\
    	\hline   
	\end{tabular}   
\end{table}

\subsection{Training Procedure}   
 
 {As is standard practice in ML, the spectroscopic redshifts (= ground truths) of each training sample are presented to the network to learn and model performance is tuned to best match the spectroscopic redshifts of the validation set. The test set is later used for an unbiased evaluation of performance, one that is independent from the entire validation tuning procedure.} 
 A simple data augmentation procedure is adopted at training time (only), whereby all images fed to the network are randomly flipped  left-right and up-down (independently). The custom data generator built for this augmented training makes sure that the photometric features are propagated unchanged to the MLP branch, while only images fed to the convnet branch are augmented.

We have experimented with training on a reduced subset of the full training data set, to characterize model performance as a function of dataset size. Our full data set comprises 136000 galaxy samples but we have also built models with only 50\%, 25\% and 12.5\% of all samples, randomly selected. For convenience, we refer to these various datasets as the 128K (full), 64K, 32K and 16K datasets. {The size of the test set remains unchanged as the training set size is varied.}

To compare validation vs. test performance with some confidence, we performed 4-fold cross-validation. This involves the compute-intensive task of training 4 times the same architecture on a different 3/4 of the training data each time, validating on the remaining 1/4 of the data.  We report validation performance as the average validation score over the 4 folds. We report test set performance as the average of the test scores produced by the 4 distinct CV-trained models on the full test dataset.
The rms variations on our CV performance scores are often less than the amount of overfitting revealed by the difference between test performance and validation performance. CV and test set performance rms variations are typically at the few \% level, rarely exceeding 5\%.  

Training was performed with an Adam optimizer, a batch size of 32 samples (fed through a custom data generator), for 200 epochs. For each CV-fold, the model with the best validation performance over 200 epochs was retained for test set performance evaluation. All model optimization choices were driven by cross-validation performance. 

We have optimized on two distinct regression objectives. In MSE-trained models, the model was optimized against the mean-squared error of predictions vs ground truth.   In MAE-trained models, the model was optimized against the mean-absolute error of predictions vs ground truth.  MAE is an $L1$-norm while MSE is an $L2$-norm. As such, MAE is known to regularize against outliers, relative to MSE (large squared errors are weighted more with MSE).

\subsection{Performance metrics}

We adopt standard performance metrics for our redshift estimation models. 
The normalized redshift bias is
\begin{equation}
\delta z = \left< \frac{z_{\rm pred} - z_{\rm true}}{1+z_{\rm true}} \right>,
\end{equation}
where $<>$ denotes the mean over data samples, $z_{\rm true}$ is the measured spectroscopic redshift and 
$z_{\rm pred}$ is the predicted redshift. We also use the $(1+z_{\rm true})$-normalized dispersion measures $\sigma$ and $\sigma_{68}$, where 
\begin{equation}
\sigma = \sqrt{ \left< \delta z^2 \right>},
\end{equation}
 is the standard deviation of the redshift bias $\delta z$ and  $\sigma_{68}$ is the sample-mean 68\% percentile spread of the absolute error on the redshift $z$ \citep[see, e.g.][for similar implementations]{2018MNRAS.475.3613S}. Finally, we report the outlier fraction beyond $3 \sigma_{68}$, referred to as $f(3\sigma_{68})$.
For comparison, ANNZ2 reports errors $\delta z \simeq 0.2-0.4 \times 10^{-3}$, $\sigma_{68} \simeq 0.03-0.05$ and $f(3\sigma_{68}) \simeq 0.04$, for their ensemble solution made of 100 distinct models built on a dataset of 180,000 SDSS DR10  galaxies \citep{2016PASP..128j4502S}.

\section{Results}

The validation set and test set performances of our deep learning solutions are presented in 
 Figures~3 and~4 for the dispersion metrics $\sigma$ and $\sigma_{68}$ and in Table~2 for the bias $\delta z$ and outlier fraction $f(3\sigma_{68})$ metrics.

\begin{figure}
	\includegraphics[width=9cm]{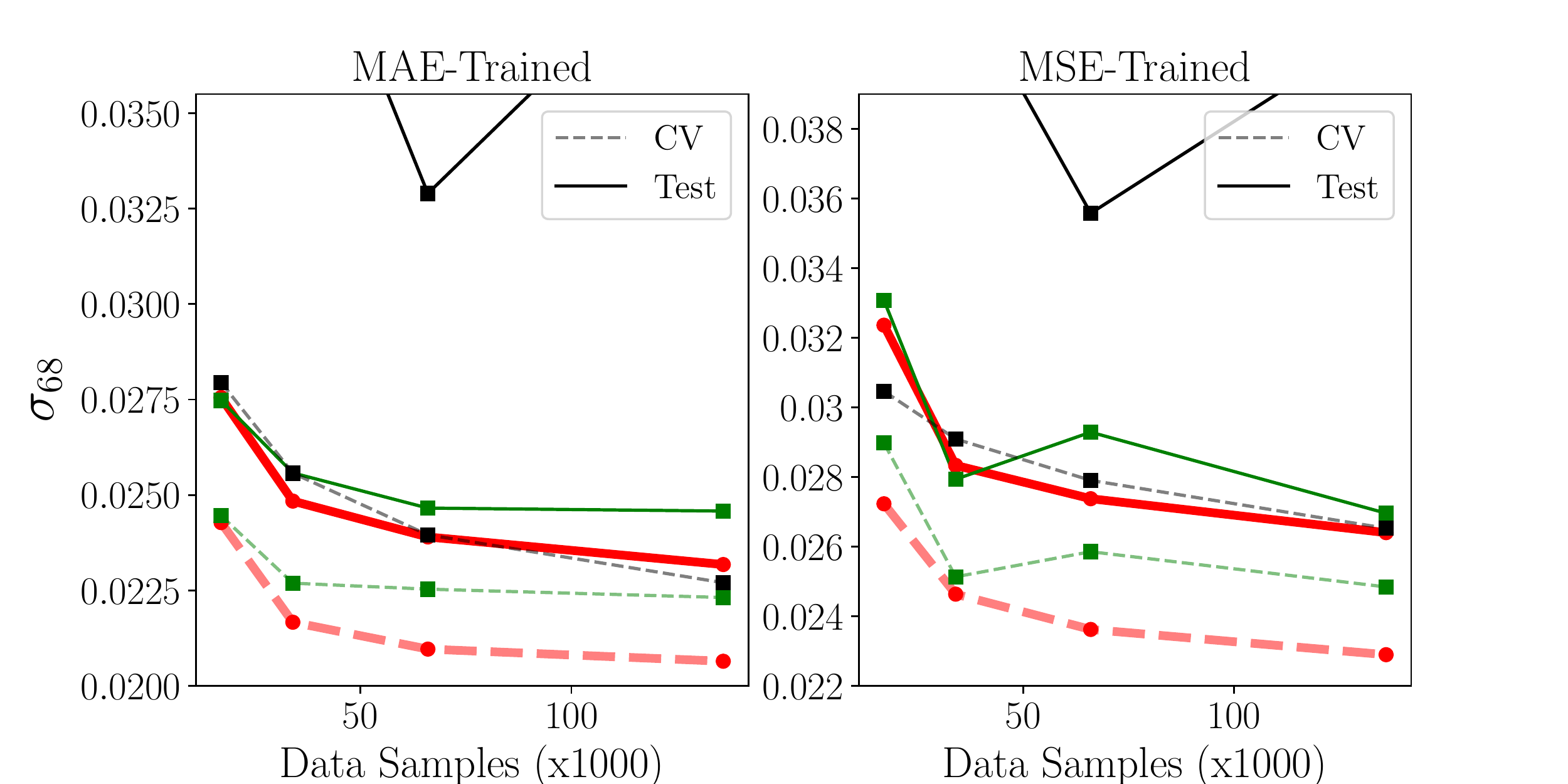}
    \caption{Performance on the $\sigma_{68}$ metric as a function of training dataset size (16K-32K-64K-128K). Models shown in the left panel are trained with an MAE (L1-norm) objective, while models in the right panel are trained with an MSE (L2-norm) objective. In each panel, solid lines show test-set performance while dashed line show cross-validation performance. Results for the mixed MLP-convnet, MLP-only and convnet-only architectures are shown in red, green and black, respectively. The best test performance is achieved by the MLP-convnet model with MAE-training, {indicating that regularization against outliers is beneficial.}}
    \label{fig:example_figure}
\end{figure}

\begin{figure}
	\includegraphics[width=9cm]{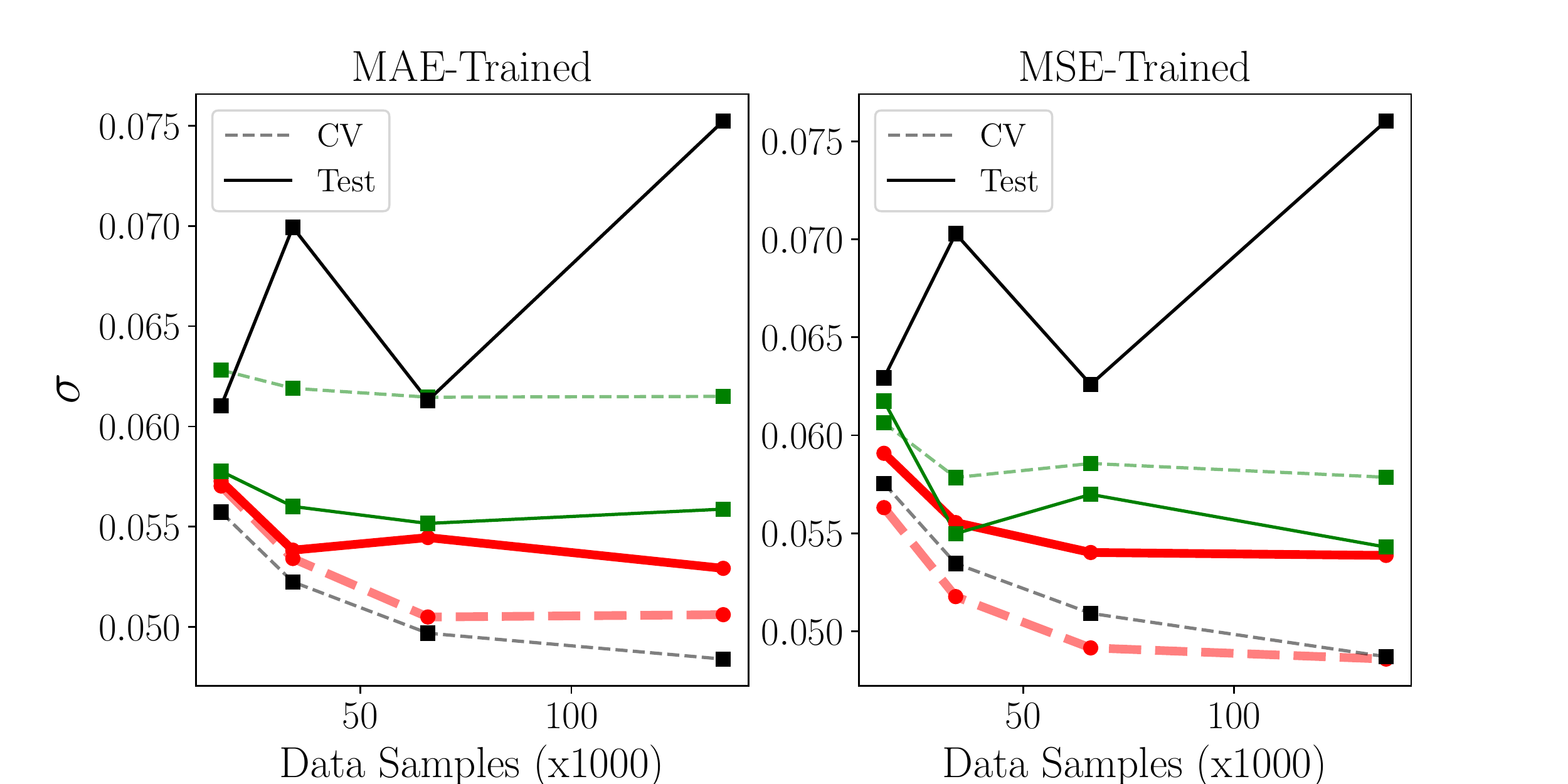}
    \caption{Same as above on the $\sigma$ metric. {The performances of MAE-trained and MSE-trained models are comparable on this metric.}}
    \label{fig:example_figure}
\end{figure}

\begin{table}
	\centering
	\caption{Convnet/MLP Test Performance: Bias and Outlier Fraction}
	\label{tab:example_table}
	\begin{tabular}{cccc} 
		\hline
		Dataset &  Model+Objective &  Bias (4 Folds) & Outlier \% \\
		\hline
		 128K    &  MLP-convnet+MAE &  $-0.70 \times 10^{-3} \pm 0.001$   			&  6.03\\
		 64K      & MLP-convnet+MAE &  	$-0.24 \times 10^{-3} \pm 0.001$			& 5.80\\
		 32K      &  MLP-convnet+MAE &    $4.26 \times 10^{-3} \pm 0.003$			& 5.94\\
		 16K      & MLP-convnet+MAE & $5.77  \times 10^{-3} \pm 0.005$				& 6.44\\
		 128K    &  MLP-convnet+MSE & $1.42  \times 10^{-3} \pm 0.002$ & 6.10 \\
		 64K     &  MLP-convnet+MSE & $0.68  \times 10^{-3} \pm 0.001$ & 5.99\\
		 32K      & MLP-convnet+MSE & $6.01 \times 10^{-3}  \pm 0.003$ & 6.41\\
		 16K      & MLP-convnet+MSE & $6.39  \times 10^{-3} \pm 0.002$ & 6.66\\
	\hline
		128K    &   MLP-only+MAE &  $50.1  \times 10^{-3} \pm 0.04$			&  2.80\\
		 64K      &   MLP-only+MAE &  	$14.3  \times 10^{-3} \pm 0.02$			& 4.43\\
		 32K      &   MLP-only+MAE &    $3.97  \times 10^{-3} \pm 0.008$ 			& 4.72\\
		 16K      &  MLP-only+MAE & $9.16  \times 10^{-3} \pm 0.05$			& 3.03\\
		 128K    &  MLP-only+MSE & $47.2  \times 10^{-3} \pm 0.04$ & 2.55 \\
		 64K     &   MLP-only+MSE & $16.1  \times 10^{-3} \pm 0.02$ & 4.80\\
		 32K      & MLP-only+MSE & $6.84  \times 10^{-3} \pm 0.01$ & 4.96\\
		 16K      &  MLP-only+MSE & $12.1  \times 10^{-3} \pm 0.05$ & 3.45\\
	\hline
		128K    &   convnet-only+MAE &  $-5.48  \times 10^{-3} \pm 0.008$			&  4.37\\
		 64K      &  convnet-only+MAE &   $18.2  \times 10^{-3} \pm 0.04$			& 4.43\\
		 32K      &  convnet-only+MAE &    $27.3  \times 10^{-3} \pm 0.03$ 			& 3.85\\
		 16K      &  convnet-only+MAE &  $-4.27  \times 10^{-3} \pm 0.008$			& 4.58\\
		 128K    &  convnet-only+MSE &  $11.1  \times 10^{-3} \pm 0.006$ & 3.97 \\
		 64K     &  convnet-only+MSE &  $18.1  \times 10^{-3} \pm 0.03$ & 2.60\\
		 32K      & convnet-only+MSE & $36.4  \times 10^{-3} \pm 0.03$ & 3.08\\
		 16K      & convnet-only+MSE & $-8.76  \times 10^{-3} \pm 0.008$ & 4.10\\

     \hline    
	\end{tabular}
\end{table}

In Figures~3 and~4, the  4-Fold CV performance (dashed lines) and test set performance (solid lines) are presented for three separate models, for various dataset sizes (16K-32K-64K-128K). The mixed MLP-convnet is shown in red, the equivalent MLP-only model is shown in green and the equivalent convnet-only model is shown in black. Figure~3 shows results on the $\sigma_{68}$ metric for training with the MAE objective (left panel) and the MSE objective (right panel). Figure~4 shows the same for the $\sigma$ metric performances. Note that the vertical scales differ in each panel of each figure.

It is worth emphasizing that the trained networks, optimized on validation performance only, are found to appreciably overfit on the validation set, which makes the test set performance crucial in being unbiased. Overfitting is particularly acute for the convnet-only model (compare the black solid and dashed lines in Figures~3 and~4). While the CV performance of the convnet-only model is decent, its test set performance is quite poor. We have not investigated the origin of this result in detail but note that it could be caused by our choice of feeding non-standard flux images (as opposed to colors) to the convnet branch of all our models. Given the generalized overfitting behavior of these models, we focus our discussion of model performance exclusively on the test set performance in what follows.

{On the $\sigma_{68}$ dispersion metrics, Figure~3 shows that the best performance is obtained when training with an MAE-objective, which regularizes against outliers. On the $\sigma$ metrics (Fig.~4), the performances of MAE-trained and MSE-trained models are comparable. } Focusing on the MAE-trained solutions, we find that the MLP-convnet model exceeds the performance of the MLP-only model. This indicates that automatically learned morphological features, once disentangled from the galaxy-integrated features fed to a separate MLP branch, do indeed improve the performance on dispersion. However, the largest gains achieved with morphology-aware models, at the level $\sim 5-10\%$ on the MAE-trained $\sigma_{68}$  and $\sigma$ metrics, remain modest.

We find that most models show some degree of improvement with the dataset size, as is generally expected for deep learning approaches \citep[e.g.,][]{DBLP:journals/corr/abs-1712-00409}.  It also worth mentioning that the MLP-only model shows some degree of underfitting on the $\sigma$ metric (solid green line below dashed green line in Figure~4). This suggests that increasing the  MLP-branch capacity (number of neurons and/or depth) might improve further model performance, a possibility that we have not explored.

The model test set performance results on the redshift bias metric, $\delta z$, and the outlier fraction metric, $f(3\sigma_{68})$,  are reported in Table~2. There is a fairly clear distinction in model performance on the redshift bias metric, with the MLP-convnet model performing better than the convnet-only and MLP-only models, especially on the larger size datasets. For comparison, ANNZ2 reports a bias metric performance $\delta z \simeq 0.2-0.4 \times 10^{-3}$ for their ensemble solution with 100 distinct models built on a dataset of 180,000 SDSS DR10  galaxies. The best performance of our MLP-convnet model (which comes from an average of 4 CV-built models) is thus approaching the performance of this ANNZ2 reference model, on a fairly comparable dataset. Outlier fractions are rather consistent across all deep learning models, although somewhat larger values are reached by the  MLP-convnet model, relative to the other two.

\subsection{Comparison to a strong baseline}

In this section, we present a strong baseline model based on well-tested machine learning algorithms and optimization strategies that do not involve deep learning, nor morphological features. Our motivation for doing so is that it is difficult to evaluate the performance of ML algorithms in isolation. Globally optimal solutions to specific ML problems are generally not known. As a result, performance comparisons, e.g. through competitions among computer scientists and machine learning practitioners, are common practice. They provide a way to rank algorithm performance consistently: the relative performances on a well-defined task can be evaluated in a controlled experiment (same dataset, same evaluation metric, etc...). 

By contrast, it is not easy (or even possible) to reasonably compare the performance of the deep learning solution presented here against that of other solutions published in the literature (because various datasets, evaluation procedures or metrics, etc... have been used). We are left with the option to present a separate baseline model, if possible with an ML methodology that differs significantly from the deep learning approach presented above. One such reference point can be obtained by building a strong baseline model with a hyperoptimized Gradient Boosting solution. Indeed, over the past few years, a specific implementation of the gradient boosting algorithm, {\tt XGBoost}, has proven very effective on numerous data science competition datasets on platforms such as Kaggle \citep{DBLP:journals/corr/ChenG16}. Together with the TPE-hyperoptimization algorithm provided in the {\tt hyperopt} module, this should constitute a strong baseline solution to the redshift estimation problem for permutation-invariant features.

\begin{table}
	\centering
	\caption{XGBoost Hyperparameter Range Explored}
	\label{tab:example_table}
	\begin{tabular}{cc} 
		\hline
		Parameter & Range \\
		\hline
		 Depth         & 4-7 \\
		 N\_estimators & 500-1750\\
		 Colsample\_bytree & 0.5-1\\
		 Subsample & 0.5-1\\
		 Learning rate & 0.01-0.2    \\                                  
     \hline    
	\end{tabular}
\end{table}

We build our baseline model with the open-source automated modeling tool {\tt MLBox},\footnote{{\tt https://mlbox.readthedocs.io/en/latest/}} which conveniently integrates key features of the {\tt scikit-learn} package, together with {\tt XGBoost} and {\tt hyperopt} built-in implementations. The model is trained on the same 10 scalar features as the MLP-branch above, described in \S2.  We optimize the 4-fold cross-validation performance of an XGBoost model over the range of hyperparameter values listed in Table~2, performing 40 hyper-optimization steps. Other XGBoost parameters are left to their default values.

As test set results show, the performance of our mixed MLP-convnet solution is comparable to the hyperoptimized XGBoost solution, which achieves best values of $\sigma_{68} \simeq 0.022$ and $\sigma \simeq 0.056$. Hyperoptimized XGboost does somewhat better on the  $\sigma_{68}$ metric but somewhat worse on the $\sigma$ metric. Note that MSE training appears generally better than MAE training for hyperoptimized XGBoost (contrary to what is found for our deep learning solution).\footnote{Another difference is that XGBoost exhibits rather minor overfit to the validation set relative to the deep learning  solutions we have presented. } On the redshift bias metric, $\delta z$, hyperoptimized XGBoost offers a consistently strong performance that is only slowly improving with dataset size ($\delta z \approx 2.5-3.0 \times 10^{-3}$, see Table~5). Our convnet-MLP solution, once trained on the largest dataset available, outperforms the XGBoost results with $\delta z  \leq 10^{-3}$ (Table~2).

Outlier fractions are fairly consistent across all the models we have built, with typical values $\sim 4-6\%$ (Tables~2 \& 5). {To facilitate comparisons with other metrics used in the photometric redshift literature \citep[e.g.][]{2017arXiv170603501C}, we also list in Table~4 values of NMAD (Normalized Median Absolute Deviation) and in Table~5 values of Outlier\_0.15 (fraction of outliers with a normalized redshift bias $\delta z >0.15$).}

Figure 5 illustrates our results with a scatter plot and marginal distributions of spectroscopic redshifts (ground truth) vs. predicted photometric redshifts from a representative XGBoost model. Specifically, Figure 5 shows 192,000 test set predictions from the hyperoptimized XGBoost model trained on the full 128K dataset with an MSE objective. This model achieves 
a bias $\delta z \approx 2.73 \times 10^{-3}$ and dispersion $\sigma = 0.056$. While bias and dispersion are small, outliers occur at a rate of a few \%, including some obvious catastrophic errors.

\begin{figure}
	\includegraphics[width=8cm]{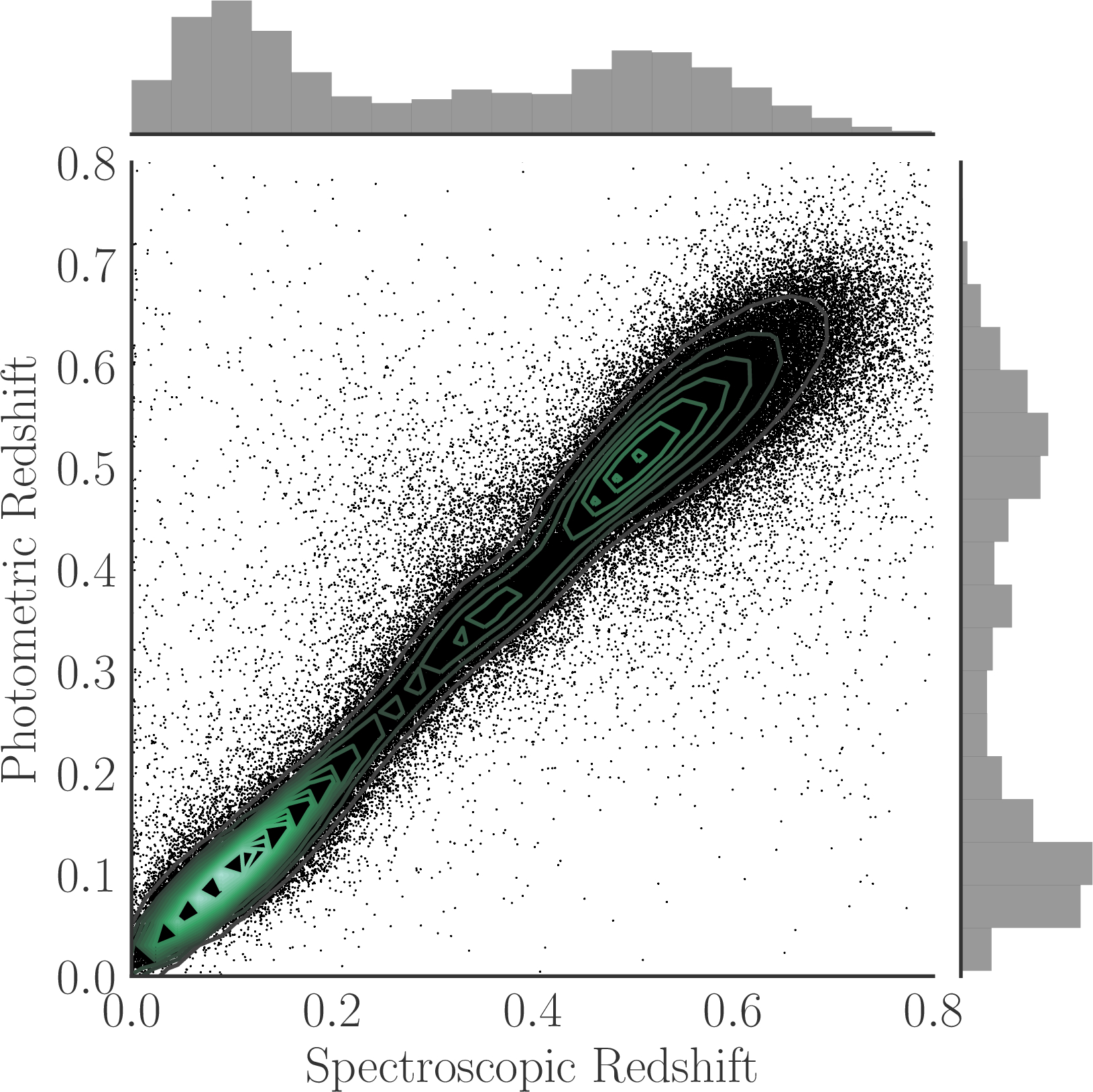}
    \caption{Representative test set performance shown as a scatter plot and marginal distributions of 192,000 spectroscopic vs. photometric redshifts predicted by a hyper-optimized XGBoost model. This specific model one was trained on the full 128K dataset with an MSE objective. The bias and dispersion are small but outliers are also easily spotted.}
    \label{fig:example_figure}
\end{figure}

\begin{table}
	\centering
	\caption{Optimized XGBoost Test Performance: $\sigma_{68}$, $\sigma$ and NMAD}
	\label{tab:example_table}
	\begin{tabular}{ccccc} 
		\hline
		Dataset & Objective & $\sigma_{68}$ &  $\sigma$ & NMAD\\
		\hline
		 128K         & MAE &  0.022 & 0.056 & 0.020 \\
		 64K         & MAE & 0.025 & 0.059 & 0.021\\
		 32K         & MAE & 0.025 & 0.059 & 0.022\\
		 16K         & MAE & 0.025 & 0.059 & 0.022\\
		 128K         & MSE & 0.022 & 0.056  & 0.021\\
		 64K         & MSE & 0.023 & 0.057 & 0.022\\
		 32K         & MSE & 0.024 & 0.057 & 0.022\\
		 16K         & MSE & 0.024 & 0.057 & 0.022\\

     \hline    
	\end{tabular}
\end{table}

\begin{table}
	\centering
	\caption{Optimized XGBoost Test Performance: Bias and Outlier Fractions}
	\label{tab:example_table}
	\begin{tabular}{ccccc} 
		\hline
		Dataset & Objective & Bias  & Outlier (\%)  & Outlier\_0.15 (\%)\\
		\hline
		 128K         & MAE &  $2.49 \times 10^{-3} $   			&  5.63 & 1.77\\
		 64K         & MAE &  	$2.52 \times 10^{-3} $			& 5.96 & 1.88\\
		 32K         & MAE &    $2.37 \times 10^{-3} $ 			& 6.14 & 2.01\\
		 16K         & MAE & $2.55 \times 10^{-3} $ 				& 6.00 & 2.01\\
		 128K         & MSE & $2.73 \times 10^{-3} $ & 5.54 &  1.68\\
		 64K         & MSE & $2.74 \times 10^{-3} $ & 5.79 & 1.76\\
		 32K         & MSE & $3.00 \times 10^{-3} $ & 5.93 & 1.87\\
		 16K         & MSE & $3.33 \times 10^{-3} $ & 5.59 & 1.85\\

     \hline    
	\end{tabular}
\end{table}

\newpage

\section{Conclusions}
We have presented a split MLP-convnet architecture aimed at helping a deep neural network disentangle strong photometric and weak morphological features. {This network outperforms the other ML solutions tested in our work on the redshift bias metric, including  the equivalent MLP-only network, which do not include morphological features.} As a result, this work establishes that morphological features indeed do help the task of photometric redshift estimation. {While the gain is small on dispersion metrics such as $\sigma$ or $\sigma_{68}$ and performance is actually worse on the outlier fraction metric, the improvement is significant on the astrophysically important redshift bias metric, $\delta z$.} Our results exploring split morpho-photometric architectures may also explain why pure convolutional approaches to redshift estimation do not necessarily outperform standard morphology-blind approaches based on color features. 

Our work has is no way exhausted the full potential of morpho-photometric redshift estimation. 
In particular, further gains in performance could be expected from various extensions of this work, such as the use of larger training sets, architecture improvements (e.g., Resnets, Densenets, etc..), additional MLP-branch optimization, additional attempts at model regularization (e.g., batch-normalization), extensions of the training process (e.g. further image augmentation such as shifts, rotations) or even model ensembling. 
In terms of applications, it would be interesting to explore whether morpho-photometric redshifts can be useful to address strong degeneracies in color space, or blending issues more frequently encountered at high redshifts, which are potentially lifted in the presence of morphological information. {It would also be beneficial to incorporate some of the probabilistic frameworks developed in the deep learning literature to obtain photometric predictions jointly with their calibrated errors.}
  
{It remains to be seen whether the gains reported with the morphology aware MLP-convnet model, in our arguably 'controlled experiment' with similar train and test set properties, will carry over to the more challenging situation of a test set with properties that deviate from those of the training set. This is a common situation for photometric redshift estimation, where a shallower spectroscopic survey is used for training, while predictions are performed on a deeper photometric survey. }

\section*{Acknowledgements}
KM is grateful to Renee Hlozek for useful discussions on the photometric redshift problem within the LSST context, as well as Ari Silburt and Dan Tamayo for comments on the manuscript. {KM thanks the referee for a report that helped improve the quality of the manuscript.} KM is supported by the National Science and Engineering Research
Council of Canada. This work has made extensive use of the following software packages: {\tt astropy}, {\tt scikit-learn},{\tt matplotlib},  {\tt seaborn}, {\tt keras}, {\tt tensorflow}, {\tt hyperopt}, {\tt MLbox}.

We would furthermore like to acknowledge that our work was performed on land traditionally inhabited by the Wendat, the Anishnaabeg, Haudenosaunee, Metis, and the Mississaugas of the New Credit First Nation.




\bibliographystyle{mnras}
\bibliography{deep_morpho} 









\bsp	
\label{lastpage}
\end{document}